\documentclass{epl}
\usepackage{epsfig}

\title{Speciational view of macroevolution: are micro and macroevolution decoupled?}

\author{V. Schw\"ammle\inst{1,3} \and E. Brigatti\inst{2}}  

\institute{
   \inst{1} Instituto de F\'{\i}sica, Universidade Federal Fluminense, 
  Campus da Praia Vermelha, 24210-340, Niter\'oi, RJ, Brazil\\
\inst{2} Centro Brasileiro de Pesquisas F\'{\i}sicas, Rua Dr. Xavier 
  Sigaud 150, 22290-180, Rio de Janeiro, RJ, Brazil\\
\inst{3} Institute for Computer Physics, University of Stuttgart, Pfaffenwaldring 27, D-70569 Stuttgart, Germany.}


\begin{document}

\maketitle

\begin{abstract}

We introduce a simple computational model that, with microscopic dynamics driven by natural
selection and mutation alone, allows the description of true speciation events. A statistical
analysis of the evolutionary tree so generated captures realistic features showing power laws
for frequency distributions in time and size. Finally, some possible interpretations of the 
absence of punctuated dynamics with mass extinctions are worked out.

\end{abstract}

The novel interpretation for old paleontological observation that Gould and 
Eldredge \cite{gould1} presented in the 70's had a deep impact on evolutionary 
theory. The theory of punctuated equilibrium leads not only to a change in the 
paradigm with which some data were analyzed, but also caused a definitive shift 
in the general way of thinking in theoretical biology. It has, in particular, 
been used as a fundamental concept to develop the idea that it is necessary to 
decouple micro and macro-evolutionary mechanisms.

Darwin's evolutionary theory understands the living world as the outcome of 
microscopic dynamics alone, driven by selection and mutation. That is, it 
supposes the existence of a causality in evolutionary changes driven by natural 
selection operating at an individual level. From this perspective, there exists 
a natural tendency to extrapolate such causality at all magnitudes and in time, 
with the hope that Darwinian natural selection alone could fully explain 
large-scale changes in history of life \cite{gould2,structure}. This 
traditional gradualistic approach focuses on how natural selection is capable 
of causing adaptation during evolution, in a process occurring at the 
population level that generates a continuous and progressive transformation of 
lineages.

In contrast, punctuated equilibrium suggests how the interposition of levels 
breaks this causal reduction and decouples micro from macroevolution. 
In this scenario, the central problem of macroevolution is to understand, 
through a direct study of species, which ones prevail and do better than the 
others, in a discrete succession of events. 

A famous model of long term analysis of evolutionary processes based on the 
idea of a decoupled macroevolution and capable of displaying a punctuated 
equilibrium behavior, is the Bak-Sneppen model \cite{bak}. Here, 
inter-specific interactions are taken into account as the predominant force
capable of generating evolution, under the simplifying assumption that the 
number of species is fixed and origination is prohibited.

In our approach, we will try to describe the whole mechanism of evolution by 
natural selection acting on individuals at a population level, in accordance 
with the most traditional Darwinistic view. Models with dynamics structured at 
the population level \cite{jorge,rikvold,lipowski,jensen} or focusing on the 
micro-macro evolution relations \cite{stauffer}, or also with a complex structure 
representing the hierarchical organization at different trophic levels 
\cite{stauffer2,cina}, are present in the literature. 

From our point of view, they are all based on interactions that can account only for the 
dynamics of extinctions. Since they characterize an individual of a new species by the appearance 
of mutation, the processes of mutation and speciation become identified and the phenomenology 
of population variance within a single species is ignored. These models are thus unable to explore some problems 
outlined by punctuated equilibrium theory because they do not implement a  
dynamical mechanism that generates speciation events within a diverse population. 

We will analyze instead a model where the 
interaction represents a natural selection responsible for speciation. Our 
purpose is to test if this driving force alone can account for all the 
phenomenology of macroevolution. For this reason, origination of new species 
is the crucial new phenomenon that our model must be able to account for. 
With this aim, we implement a self-modifying selective force based 
on frequency dependent selection \cite{dieckmann,edgardo} that allows 
coexistence and branching of taxa. To sum up, we do not consider species-level 
fitness, but a mechanism that generates species autonomously. 
We do not simply perform a refilling of extinct species (as for example in 
Refs.\cite{bak,jorge,stauffer,sole1}) and, as a consequence, their number is 
not fixed. 
This approach also unifies the three time scales \cite{maynard} 
that characterize evolution: the fast population dynamical scale (controlled by natural
selection), the slow evolutionary scale (controlled by the mutation  
process) and the ultra slow macroevolution (the timescale of the 
speciation/extinction dialectic). Although there is no doubt that single 
speciation and extinction events occur by the interaction between natural 
selection and mutation (slow and fast scales), as stated above, there is no 
general agreement whether macroevolution can be seen as the simple 
consequence of the speciation events generated by population dynamics, 
without the necessity of accounting for interactions at other levels. 

We will face this question by comparing the results of our model, born from 
this unifying view of evolution, with all the quantitative statistical 
properties observed in the fossil record: scale free behavior for at least 
some range of the distribution in time and size \cite{sepkoski} and a time 
series of extinction events showing punctuated equilibrium \cite{gould2} where 
we can find mass extinctions \cite{raup} and long term correlations
\cite{sole3} predominate. We will see that, although the scale free nature is 
reproduced, not all characteristics linked with punctuated equilibrium show up 
in the results. 

For reasons of simplicity, the model is not developed in genotype space, as for 
instance in Refs.\cite{jorge,rikvold,jensen}, but in the more easy-to-handle strategy 
space where an individual is represented by an integer number, the strategy 
parameter $x$ ($0 \le x \le P$), that takes into account all the phenotypic 
characteristics that determine its biological success. At each time step, an 
individual generates one offspring with the same strategy as its parent, 
eventually mutated by a random $\pm 1$ factor with some probability $\mu$,
that is kept constant from the start of the simulation. 

We allow each agent to live until the occurrence of death caused by a selective 
pressure. This natural selection is characterized by two different components. 
The first is density-dependent, responsible for limiting the size of the total 
population and controlled by the carrying capacity. The other, a 
frequency-dependent factor, takes into account how, in realistic situations, 
the tendency to occupy a more favored region in strategy space balances with an 
increasing competition among individuals. The latter is the dynamic component 
of selection, which represents the feedback between individuals and ecosystem 
and takes into account the instantaneous distribution of the population. Thus, 
natural selection is implemented through a death probability that, in the Monte 
Carlo simulation, takes the form: 
\begin{eqnarray}
S = \frac{1}{K} \cdot \sum \limits_{y=0}^{P} N_{y}\cdot\exp(-\frac{(x-y)^2}{2b^2})
\end{eqnarray} 
At each time step, a random number is tossed; the individual survives if this 
number is  larger than $S$. The strength of competition declines with 
distance in strategy space according to a Gaussian function with deviation 
$b$, and parameter $K$ depicts the carrying capacity. $x$ is the value of the 
strategy parameter of the individual that is feeling the selection pressure, 
and the sum runs over the $y$ index that spans all of strategy space. 
By $N_{y}$ we indicate the number of individuals with strategy $y$.
We use periodic boundary conditions in order to avoid edge effects.
This selective function, inspired by the ones in Refs.\cite{dieckmann,edgardo}, 
represents non-local interactions between individuals and introduces a finite wave-length
instability. 
This selection is repulsive for individuals living in crowded
regions of x-space, allowing branching and/or extinction.
The mutation rate allows the conquest of new areas in strategy
space and thus generates the fluctuations that can lead to the 
self-organization of a varying number of different strategy clusters. 
For this reason, it does not drive 
the system to an optimal ending point, but leaves it in a permanently 
changing dynamic state.

Since we deal with an asexual population, the biological characterization of 
species, defined for sexual individuals as a reproductively isolated 
population, must be substituted by a more operational definition, based on a 
functional differentiation among phenotypical distinct groups. For this 
reason, we refer to species as a group of individuals that share most of 
their phenotypic features but which differ in a few traits. According to 
this definition, the algorithm used associates different species to different 
clusters of individuals that have a small strategy distance - one being 
already enough - between them. That is, the space between two clusters can 
not be occupied by individuals. Although spatial heterogeneity or predation 
may have a relevant influence on the dynamics of the population and, in 
particular, on the frequency of branching events, they are not taken into 
account in our model. Additionally, a static selection component that 
defines the general ecological condition and can cause a directional 
selection, driving the population towards some fitness maximum in strategy 
space, does not change crucially the dynamics of the model.

\begin{figure}[p]
\begin{center}
\vspace*{0.8cm}
\centerline{\psfig{figure=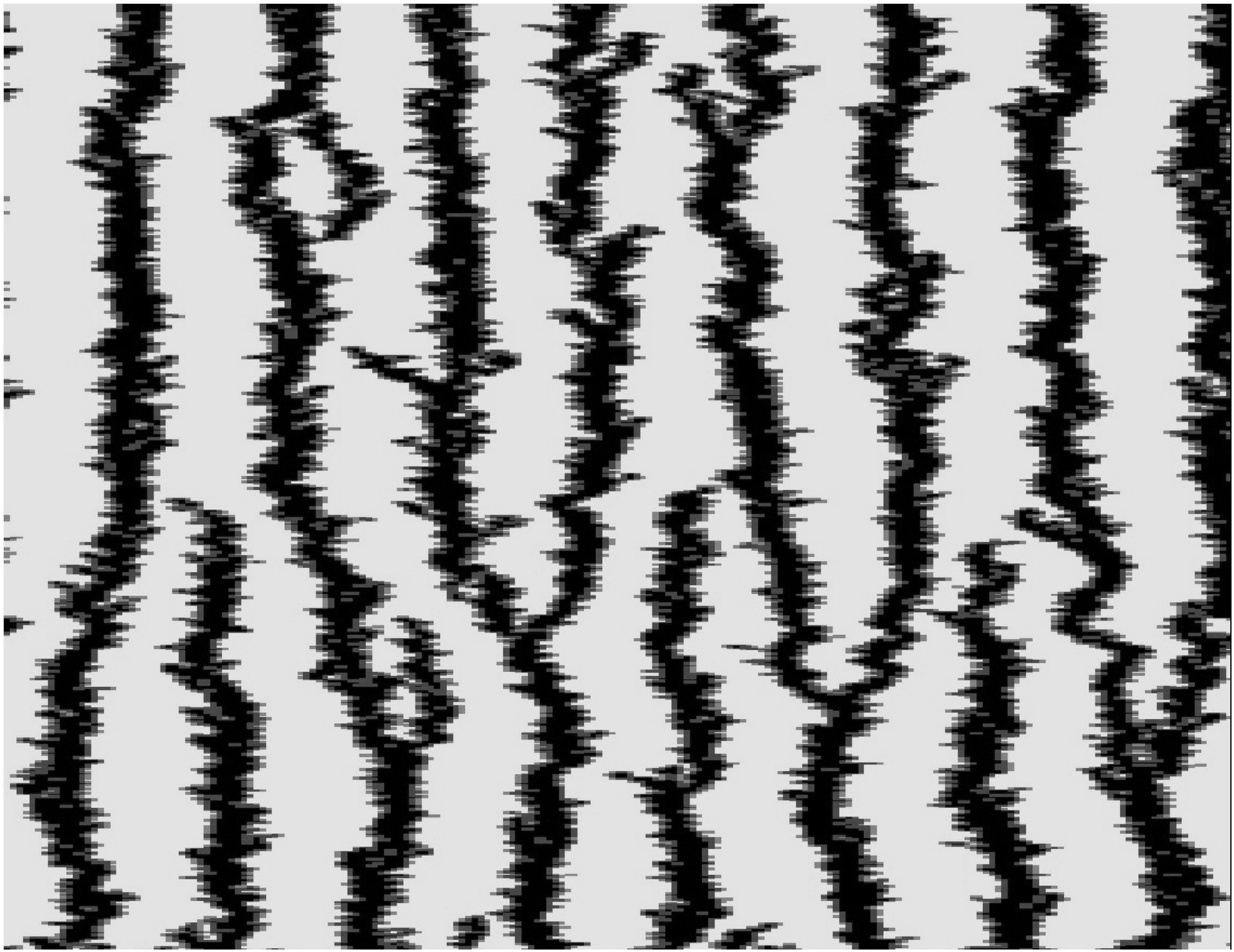, width=8.5cm, angle=0}}
\vspace*{0.4cm}
\end{center}
\caption{ Time evolution of the population: the horizontal axis represents the strategy 
space ($150<P<300$), the vertical one the time ($200000$ time steps interval). 
The simulation ($K=10000$, $\mu=0.005$, $P=500$ and 
$b=15$) started with all the strategy space filled by a uniform distribution. 
Anyway, other initial conditions (such as, for example, a single
species) generate, after a short transient, evolutionary trees that can not be distinguished 
from one another.}
\label{fig_tree}
\end{figure}

The dependence of the model's behavior on the value of the parameters can be 
summarized by some simple rules. In general, the parameters of interest are 
only two, and are the ones that effectively control the branching 
probability. The carrying capacity ($K$) and the number of possible strategies 
($P$), on the other hand, are not so crucial. In fact, the only role of the 
first is to regulate the population size, while $P$ is correlated with the 
mean number of living species, which grows with $P$ following a linear 
relation. In contrast, the mutation rate ($\mu$) directly influences the 
branching probability controlling the strategy variability of the population 
and both the broadness of the distribution and the speed of the process. In the 
following we fix the value of $\mu$ so as to make possible a realistic 
evolutionary simulation, where mutations have to occur infrequently, which is 
in accordance with the fact that this parameter controls the dynamics on the 
slow evolutionary scale. For this reason, we set its value to $0.005$. One 
parameter remains, $b$, whose value is responsible for controlling competition 
and, as a consequence, the force that splits up a cluster into two different 
ones. This drive increases as $b$ is decreased, causing a larger number of 
occurrences of branching events. There is a simple relation between the mean 
number of species and $b$, taking the form: $N(b) \propto b^{-1}$. We adjust 
this last parameter by searching for an equilibrium between really slow 
branching dynamics, which happens for large $b$ and is a case not suitable for a 
statistical analysis, and small $b$ values, for which the population feels 
such a strong drive that it is impossible to define an evolutionary 
tree. In this last situation, where the branching events are so numerous that 
the distribution can not be well defined, with a large number of peaks 
connected by intermediate strategies, it is impossible to perform a cluster 
analysis. An example of a realistic and living evolutionary tree generated for 
standard parameters values can be seen in fig.~\ref{fig_tree}.

\begin{figure}[p]
\begin{center}
\vspace*{0.8cm}
\centerline{\psfig{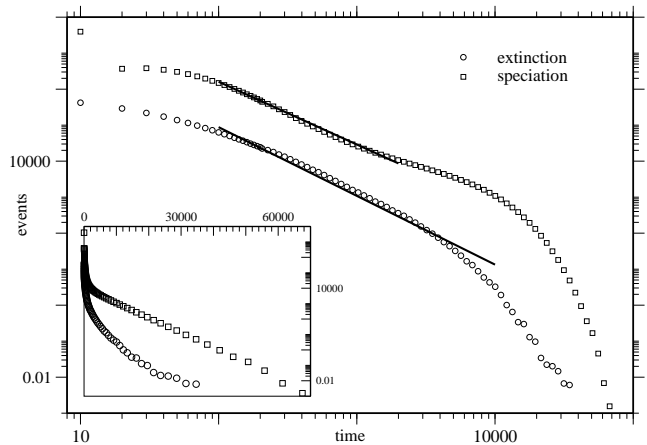}}
\vspace*{0.4cm}
\end{center}
\caption{Frequency distribution for species' lifetimes. The circles represent 
the lifetime of extinction events (from the branching of the new taxon until 
its extinction), the squares the lifetime between speciation events. The inset 
is an evidence of the exponential tails at large times. The simulation run had 
the duration of $10^7$ time steps and its parameters, used also in the 
simulations that follow, were: $K=10000$, $\mu=0.005$, $P=500$ and $b=15$.}
\label{fig_life}
\end{figure}

We start our analysis investigating the probability distribution of lifetimes 
$E(t)$ of the species, a central measure due to its comparability with 
observational results. From the data shown in fig.~\ref{fig_life}, we observe 
that a power law can be recovered
\begin{eqnarray}
E(t) \propto t^{-\gamma}\,\,\, ; \,\,\, \gamma = 2.02 \pm 0.04
\end{eqnarray}
over about two decades, with an exponential tail for large times. 
These values are comparable to data from extinction records. Even if their 
interpretation is still under debate, it seems that a power law fitting with 
an exponent close to $-2$ is more convincing than an exponential, 
for at least the shorter lifetimes  \cite{sepkoski,drossel}. 
A similar behavior was confirmed by other models as well: 
Refs.\cite{jorge,rikvold,jensen} agree with our value for 
the exponent and Refs. \cite{stauffer,stauffersm} with the deviation from the 
power law for very long lifetimes.

The same figure shows also data related to the distribution of lifetimes for 
the origination processes. The life time of originations represents the time 
interval between one speciation event and the following one in the same 
lineage. These new data are of more difficult interpretation because there 
are no observations in the fossil records. Moreover, only models where 
speciation events do not coincide with extinction events (such as they do in 
refilling models) but are defined by an internal dynamics can produce such 
results. The distribution of lifetimes of originations shows also a power law 
behavior with $\gamma = 1.64 \pm 0.01$, with a rather extended exponential tail.

\begin{figure}[p]
\begin{center}
\vspace*{0.8cm}
\centerline{\psfig{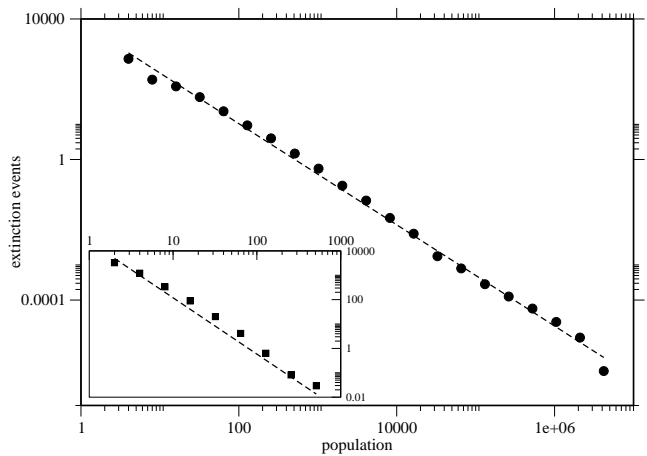}}
\vspace*{0.4cm}
\end{center}
\caption{Frequency distribution for extinction size. We counted the number of 
all the individuals, from the branching of the new species until its extinction.
The inset shows the number of individuals related to a taxon, normalized for 
its lifetime. In this case, the power-law takes the form: 
$E(s) \propto s^{ -2.19 \pm 0.07}$.}
\label{fig_dim}
\end{figure}

From our simulations we also obtain the distribution of extinction events as 
a function of their size $s$. By the term size we denote the number of 
individuals that make part of an extinct taxon from its origination until its 
disappearance (see fig.~\ref{fig_dim}). It is possible to fit the data with 
a power-law:   $E(s) \propto s^{ -1.44 \pm 0.03}$. It is difficult to compare  
these results with the other ones present in the literature 
\cite{jorge,sole1} because usually the size of the events is obtained by 
counting the number of species or families.

\begin{figure}[p]
\begin{center}
\vspace*{0.8cm}
\centerline{\psfig{figure=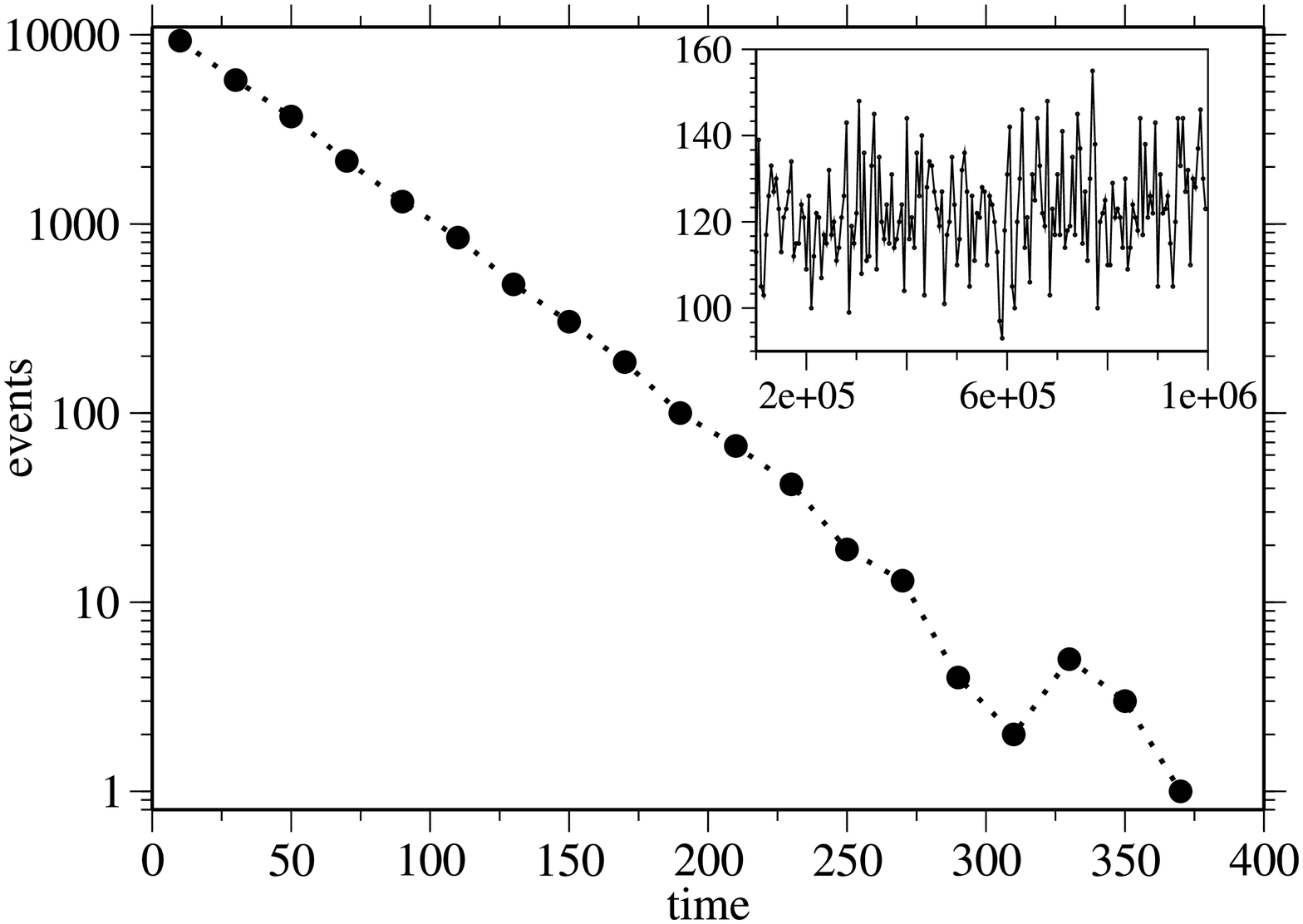, width=8.5cm, angle=0}}
\vspace*{0.4cm}
\end{center}
\caption{Exponential distribution of the periods of stasis. The inset shows the 
temporal evolution of the number of extinction events, as obtained by 
collecting each value in a time interval of $5000$ steps.}
\label{fig_stasis}
\end{figure}

From the results stated above it seems that our model is characterized by 
power law distributions in time and size, though restricted to some decades. 
However, by analyzing the distribution of lengths of 
intervals without activity (period of stasis), we find a clear exponential 
behavior (see fig.~\ref{fig_stasis}). If a critical process was involved, we 
should expect another power law. Moreover, the existence of a scale, that 
breaks any possible continuous connection between small, intermediate and large 
extinctions, is easily perceived by analyzing the time evolution of the number 
of extinctions (fig.\ref{fig_stasis}). No mass extinctions are present and in 
the time series we can not recover an intermittent behavior characteristic of 
a punctuated equilibrium phenomenon. 

Finally, we tried to detect long-range correlations in the time series of 
extinction events (inset of fig.~\ref{fig_stasis}). For correlated events, 
a fluctuation $F \propto t^{\alpha}$ with $\alpha \ne 1/2$ is expected. 
Moreover, the exponent $\alpha$ is related to the one ($\beta$) describing the 
power spectrum of the series through: $2\alpha = \beta +1$.  Although this
result is still controversial, the 
study of some paleontological data \cite{sole3} suggested self-similar 
fluctuations described by an $1/f$ spectrum ($\beta =1$). 
We analysed some time series (differing  in the range of 
the sampling interval or in the $\mu$ value) using the DFA 
method \cite{DFA}. No correlations were found ($\alpha = 1/2$). \\

Our model presents promising results showing, in accordance with 
observations gathered from the fossil records, a power law behavior for 
statistical distributions in time and size. Albeit these successful 
predictions, the last results show a difficulty in obtaining punctuated 
dynamics with mass extinctions, where long range correlations allow the 
clustering of the extinction events. These facts are somewhat intriguing and 
deserve further investigation. 

It is true that the behaviour shown in our data may be an artifact of the structure of 
our model, which is based on an autoregulating selection that does not allow for large 
fluctuations in the population size.  Even if this is the case, we can look at these results 
as a contribution to the discussion about the possibility for a unified model to represent 
long-term statistics of evolution.

Our model has a microdynamics of speciation based only on 
a natural selection force, which operates on organisms and is the exclusive 
responsible for adaptive evolutionary changes, in accordance with the original 
Darwinian paradigm \cite{structure}.
We are aware that our model is just one possible implementation of such a scenario 
(selection based on competition) and is, in general terms,
highly simplified. For these reasons it is not capable of supporting a definitive claim.
Nevertheless, it is a step in the direction of stimulating the analysis in more specific frames. 
Some important questions are risen by our work: (a) is it sufficient to implement another more realistic 
and richer microevolutionary mechanism or (b) is it necessary to take into account 
other macroevolutionary dynamics added to natural selection \cite{gould1} or, 
finally,  (c) could it be enough to include external stresses, related with mass 
extinction events alone \cite{newman}?

Whatever the right answer is, we can claim to have been able to describe, 
through the implementation of true speciation events, the statistical distributions 
related to the spontaneous rate of replacement of one species by another.

\acknowledgements
Both authors contributed equally to this work. V.S. is funded by the DAAD 
(Deutscher Akademischer Austauschdienst); E.B. was supported by the Brazilian 
Agency CAPES and CNPq. We thank J.S. S\'a Martins and S. Moss de Oliveira 
for a critical reading of the manuscript.

\end{document}